\documentstyle[12pt]{article}
\newcommand{\beq}{\begin{eqnarray}}
\newcommand{\eeq}{\end{eqnarray}}
\newcommand{\k}{{\bf{k}}}
\newcommand{\r}{{\bf{r}}}
\newcommand{\p}{{\bf{p}}}

\newcommand{\F}{{\bf{F}}}
\newcommand{\om}{{\omega}}
\begin{document}
\thispagestyle{empty}
\title
{
L\'{e}vy flights in random environments
}
\author
{
Hans C. Fogedby$^{\dag}$\\
Institute of Physics and Astronomy\\
University of Aarhus\\
8000 Aarhus C\\
Denmark
}
\maketitle
\begin{abstract}
We consider L\'{e}vy flights characterized by the step
index $f$ in a quenched
random force field. By means of a dynamic renormalization group
analysis we find that the dynamic exponent $z$ for $f<2$ locks onto
$f$, independent of dimension and {\em independent}
of the presence of weak quenched disorder.
The critical dimension, however, depends on the
step index $f$ for $f<2$ and is given by $d_c=2f-2$.
For $d<d_c$ the disorder
is {\em relevant}, corresponding to a non trivial fixed point
for the force correlation function.
\end{abstract}


[$\dag$] E-mail: fogedby@dfi.aau.dk
\newpage

There is a current interest in the dynamics of fluctuating manifolds
in quenched random environments \cite{1}. This fundamental issue
in modern condensed matter physics is encountered in problems as
diverse as vortex motion in high temperature superconductors,
moving interfaces in porous media, and random
field magnets and spin glasses.

The simplest case is that of a random walker in a random environment,
corresponding to a zero dimensional fluctuating manifold. This
problem has been treated extensively in the literature
\cite{alex1,havlin,bouch1} and many results are known.

In the case of ordinary Brownian motion, characterized by a finite mean
square step, in a pure environment without disorder, the central limit
theorem \cite{feller} implies that the statistics of the walk
is given by a Gaussian
distribution with a mean square deviation proportional to the number
of steps or, equivalently, the elapsed time, i.e., the mean square
displacement
\beq
\langle r^2(t)\rangle\,\propto\, Dt^{2/z},
\eeq
where the dynamic exponent $z=2$ for Brownian walk and $D$ is the
diffusion coefficient.

There are, however, many interesting processes in nature which are
characterized by an anomalous diffusion with $z\neq 2$ due to the
statistical properties of the environments \cite{havlin,bouch1}.
Examples are found in chaotic systems \cite{geisel}, turbulence
\cite{grossman,bohr}, flow in fractal geometries \cite{sahimi}, and
L\'{e}vy flights \cite{shlesinger1,fogedby}, which generally lead
to enhanced diffusion or superdiffusion with $z<2$. We note that
the ballistic case corresponds to $z=1$. The other case of
subdiffusion or dispersive behaviour with $z>2$ is encountered
in various constrained systems like doped crystals, glasses or
fractals \cite{scher,shlesinger2,alex2,blumen1,blumen2}.

Irrespective of the spatial dimension $d$, ordinary Brownian
motion traces out a manifold of fractal dimension $d_F=2$
\cite{hughes}.
In the
presence of a quenched disordered force field in $d$ dimensions the
Brownian walk is unaffected for $d>d_F$, i.e., for $d$ larger than
the critical dimension $d_c=d_F$ the walk is transparent and the
dynamic exponent
$z$ locks onto the value $2$ for the pure case.
Below the critical dimension $d_c=2$ the long time characteristics
of the walk is changed to subdiffusive behaviour with $z>2$
\cite{fisher1,der,luck}. In $d=1$, $\langle r^2(t)\rangle\propto [\log{t}]^4$,
independent of the strength of the quenched disorder \cite{sinai}.

L\'{e}vy flights constitute an interesting generalization of ordinary
Brownian walks.
Here the step size is drawn from a L\'{e}vy distribution
characterized by the step index $f$ \cite{feller}. The L\'{e}vy distribution
has a long range algebraic tail corresponding to large but infrequent
steps, so-called {\em rare events}. This step distribution has the interesting
property that the central limit theorem does not hold in its usual
form. For $f>2$ the second moment or mean square deviation of the step
distribution is finite, the central limit theorem holds and the dynamic
exponent $z$ for the L\'{e}vy walk locks onto $2$, corresponding to ordinary
diffusive behaviour; however, for $f<2$ the mean square step deviation
diverges, the rare large step events prevail and determine the
long time behaviour, and the the dynamic exponent $z$ depends on
the microscopic step index $f$ according to $z=f~~(f<2)$, indicating
anomalous enhanced diffusion, that is superdiffusion
\cite{fogedby,klafter,blumen3}. The `built in' superdiffusive
characteristics of L\'{e}vy flights have been used to model a variety
of physical processes such as
self diffusion in micelle systems \cite{ott}, and transport
in heterogeneous rocks \cite{klafter}.

In the present letter we consider L\'{e}vy flights in the presence of a
quenched random force field and examine the interplay between the
`built in' superdiffusive behaviour of the L\'{e}vy flights and the
pinning effect of the random environment generally leading to
subdiffusive behaviour. Generalizing the discussion in
refs.\cite{fisher1,luck,aronovitz} we find that in the case of
enhanced diffusion for $f<2$ the dynamic exponent $z=f$, independent
of the presence of weak disorder. On the other hand, we can still
identify a critical dimension $d_c=2f-2$, depending on the step
index $f$ for $f<2$.
Below $d_c$ the quenched
disorder becomes {\em relevant} as indicated by the emergence of a
non trivial
fixed point in the renormalization group analysis. Here we give
a brief outline of our arguments while details will be published
elsewhere.

It is convenient to discuss L\'{e}vy flights in terms of a Langevin
equation with `power law' noise \cite{zhang}. In an arbitrary drift force
field $\F(\r)$, representing the quenched disordered environment,
the equation takes the form
\beq
\frac{d\r(t)}{dt}=\F(\r(t))+\mbox{\boldmath$\eta$}(t).
\eeq
Here \mbox{\boldmath$\eta$} is the instantly correlated power law white
noise with distribution
\beq
p(\mbox{\boldmath$\eta$})d^d\eta\propto\eta^{-1-f}d\eta.
\eeq
We assume an isotropic distribution characterized by the step index
$f$. In order to ensure normalizability we have introduced a
lower cutoff
$\eta\sim a$ of the order of a microscipic length $a$ and chosen $f>0$. For
$f>2$ the second moment, $\langle\eta^2\rangle=
\int p(\mbox{\boldmath$\eta$})\eta^2d^d\eta$, is finite and
a characteristic step size is given by the root mean square deviation
$\sqrt{\langle\eta^2\rangle}$. For $1<f<2$ the second moment
diverges but the mean step,
$\langle\eta\rangle$, is finite. In the interval
$0<f<1$ the first moment diverges and even a mean step size is not defined.

The power law noise $\eta$, describing the consecutive L\'{e}vy steps,
drives the position $\r$ of the walker. For the quenched force field
$\F(\r)$ we assume a Gaussian distribution,
\beq
p(\F)\propto\exp{(-\frac{1}{2}
\int d^drd^dr'
F^\alpha(\r)
\Delta^{\alpha\beta}(\r-\r')^{-1}
F^\beta(\r'))}
\eeq
\beq
\langle F^\alpha(\r)F^\beta(\r')\rangle_F=\Delta^{\alpha\beta}(\r-\r').
\eeq
Here $\Delta^{\alpha\beta}(\r-\r')$ is the force correlation
function expressing the range and vector nature of $\F(\r)$.

In the absence of the force field, i.e., $\F(\r)=0$, the
distribution for the position of the walker is easily inferred.
{}From the definition $P(\r,t)=<\delta(\r-\r(t))>$, the solution
of Eq. (2), and averaging according to Eq. (3), we find the scaling
form \cite{fogedby,hughes},
\beq
P(\r,t)=\int\frac{d^dk}{(2\pi)^d}
\exp{(i\k\r-Dk^\mu\mid t\mid)}=
\mid t\mid^{-\frac{d}{\mu}}G(r/\mid t\mid^{\frac{1}{\mu}}).
\eeq
$D$ is a diffusion coefficient setting the time scale.
This procedure is equivalent to applying the central limit
theorem to a sum of random variables with the L\'{e}vy
distribution in Eq. (3) \cite{feller,hughes}.

The scaling exponent $\mu$ depends on the step index $f$,
characterizing the L\'{e}vy distribution. For $f>2$, i.e., the case
of a finite mean square step, $\mu$ locks onto the value $2$ and
the scaling function $G$ takes the  Gaussian form for ordinary
Brownian walk, $G(x)=\exp{(-x^2)}$. This
is a consequence of the central limit theorem which here
leads to universal behaviour. For $f<2$ the scaling exponent
$\mu=f$ and the scaling function $G$ can only be given explicitly
in terms of known functions for $\mu=1$, the ballistic case, where
we find the Cauchy distribution $F(x)=(1+x^2)^{-((d+1)/2)}$. It is,
however, easy to show that $G\rightarrow const.$ for $x\rightarrow 0$
and $G\rightarrow 0$ for $x\rightarrow\infty$.
{}From the distribution in Eq. (6)
we deduce the scaling form for the mean square displacement
of the walker in time $t$:
\beq
<r^2(t)>=\int P(\r,t)r^2d^dr
\propto
t^{\frac{2}{\mu}}
\eeq
and following from Eq. (1) the dynamic exponent $z=\mu$.
We also note that the fractal dimension of a L\'{e}vy flight
is $d_F=\mu$ \cite{hughes}.

In the presence of the quenched force field given by
Eqs. (4) and (5) it is convenient to recast the problem given
by the Langevin equation (2) in terms of the associated
Fokker-Planck equation,
\beq
\frac{\partial P(\r,t)}{\partial t}=
-\nabla(\F(\r)P(\r,t))+D\nabla^\mu P(\r,t).
\eeq
Here the first term on the right hand side of Eq. (8) is the usual
drift term due to the motion of the walker in the force field,
the second term follows from Eq. (6) and the assumption of
independent contributions to the probability current. The
`fractional' gradient operator $\nabla^\mu$ is the Fourier
transform of $-k^\mu$ and is a spatially non local integral operator
reflecting the long range L\'{e}vy steps; for $\mu =2$ it reduces
to the usual Laplace operator describing ordinary diffusion.

The Fokker-Planck equation in Eq. (8) together with the distribution
in Eqs. (4) and (5) defines the problem of a random L\'{e}vy
walker in a quenched random force field. The `microscopic'
L\'{e}vy steps occuring on a fast time scale represented by the
noise term in the Langevin equation (2) has now been absorbed
and replaced by the anomalous diffusion term in the Fokker-Planck
equation (8). The remaining randomness due to the quenched
force field is assumed static compared to the time scale of
$\r(t)$.

There are a variety of techniques available in order to treat the
random Fokker-Planck equation (8). Applying the Martin-Siggia-Rose
formalism in functional form \cite{martin,dedom1,dedom2,bauch}
and using either the replica method \cite{bouch1} or an explicit
causal time dependence \cite{fisher1,dedom2,pel}, one can average
over the quenched force field and construct an effective field
theory. A more direct method, which we shall adhere to in the
present discussion, amounts to an expansion of the Fokker-Planck
equation (8) in powers of the force field and an average over
products of $\F(\r)$ according to the distribution in
Eqs. (4) and (5) \cite{aronovitz}.

In order to deduce the scaling properties of the force averaged
distribution \linebreak $\langle P(\r,t)\rangle_F$
and the mean square displacement
$\langle\langle r^2(t)\rangle\rangle_F$ we carry out a renormalization group
analysis of
Eq. (8) following the momentum shell integration method propounded
in refs. \cite{forster1,forster2,aronovitz} and also drawing
on the results in ref.\cite{fisher1}.

Defining the Fourier transform
\beq
P(\k,\om)=\int d^drdt\exp{(i\om t-i\k\r)}
P(\r,t)\theta(t),
\eeq
where $\theta(t)$ is the step function, we obtain the Fokker-Planck
equation
\beq
(-i\om+Dk^\mu)P(\k,\om)=P_o(\k)-
i\k\int\frac{d^dp}{(2\pi)^d}\F(\k-\p)P(\p,\om).
\eeq
Here the force field is averaged according to Eqs. (4) and (5), or
\beq
\langle F^\alpha(\k)F^\beta(\p)\rangle_F=
\Delta^{\alpha\beta}(\k)(2\pi)^d\delta(\k+\p)
\eeq
for all pairwise force contractions (Wick's theorem for the
Gaussian distribution);
$P_o(\k)=P(\,t=0)$ is the initial distribution.
For simplicity we consider the case of isotropic zero range force
correlations, i.e., $\Delta^{\alpha\beta}(\k)=\Delta\delta^{\alpha\beta}$;
the general case has been discussed in
refs.\cite{bouch1,aronovitz,fisher2}. We introduce a microscopic
UV cut off and assume $0<k,p<1$. Iterating Eq. (11), identifying
self energy and force correlation corrections to second order
in $\Delta$, and performing a momentum shell integration, i.e.,
averaging over the force in the shell $e^{-\ell}<k,p<1$,
we obtain the `corrected' Fokker-Planck equation
\beq
(-i\om+Dk^\mu+\delta Dk^2)P(\k,\om)=P_o(\k)-
i\k\int\frac{d^dp}{(2\pi)^d}\F(\k-\p)P(\p,\om).
\eeq
and the force correlation function
\beq
\langle F^\alpha(\k)F^\beta(\p)\rangle_F=
(\Delta+\delta\Delta)\delta^{\alpha\beta}(2\pi)^d\delta(\k+\p)
\eeq
for $0<k,p<e^{-\ell}$. For small values of the scale parameter
$\ell$ the corrections $\delta D$ and $\delta\Delta$ arising from
the momentum shell integrations are proportional to $\ell$. From
the diagrammatic contributions to $\delta D$ and $\delta\Delta$
given in refs. \cite{fisher1,aronovitz}, involving three
propagators $(-i\om+Dk^\mu)^{-1}$ and two force contractions
$\Delta$ for $\delta D$ and two propagators and two force
contractions for $\delta\Delta$, all evaluated in the static limit
$\om=0$ and on the shell $k=1$, we obtain
\beq
\delta D=-A\frac{\Delta^2}{D^3}\ell~~~~~~~
\delta\Delta=-B\frac{\Delta^2}{D^2}\ell,
\eeq
where $A$ and $B$ are geometric factors associated with the area
of a d-dimensional unit sphere. In order to derive
the `renormalized' Fokker-Planck equation we introduce
scaled quantities
$\k^{\prime}=\k e^{\ell}$,
$\p^{\prime}=\p e^{\ell}$,
$\om^{\prime}=\om e^{\alpha(\ell)}$,
$P^{\prime}(\k^{\prime},\om^{\prime})=P(\k,\om)e^{-\alpha(\ell)}$,
and
$\F^{\prime}(\k^{\prime})=\F(\k)e^{-(d+1)\ell+\alpha(\ell)}$
such that $0<k^{\prime},p^{\prime}<1$ and find
\beq
& &(-i\om^{\prime}+Dk^{\prime\mu}e^{-\mu\ell+\alpha(\ell)}
+\delta Dk^{\prime 2}e^{-2\ell+\alpha(\ell)})
P^{\prime}(\k^{\prime},\om^{\prime})=\nonumber\\
& &P_o^{\prime}(\k^{\prime})-
i\k^{\prime}\int\frac{d^dp^{\prime}}{(2\pi)^d}
\F^{\prime}(\k^{\prime}-\p^{\prime})P^{\prime}(\p^{\prime},\om^{\prime})
\eeq
and
\beq
\langle F^{\prime\alpha}(\k^{\prime})F^{\prime\beta}(\p^{\prime})
\rangle_F\,=\,
(\Delta+\delta\Delta)e^{-(d+2)\ell+2\alpha(\ell)}
\delta^{\alpha\beta}(2\pi)^d\delta^d(\k^{\prime}+\p^{\prime}).
\eeq
{}From the renormalized Fokker-Planck equation and force
correlation function we read off the renormalization group
equations for $D$ and $\Delta$,
\beq
D^{\prime}=De^{\mu\ell+\alpha(\ell)}
\eeq
\beq
\Delta^{\prime}=(\Delta+\delta\Delta)e^{-(d+2)\ell+2\alpha(\ell)}
\eeq
or in differential form setting
$\alpha(\ell)=\int_o^\ell z(\ell^{\prime})d\ell^{\prime}$
and introducing Eq. (14)
\beq
\frac{dD(\ell)}{d\ell}=(z(\ell)-\mu)D(\ell)
\eeq
\beq
\frac{d\Delta(\ell)}{d\ell}=(2z(\ell)-d-2)\Delta(\ell)
-B\frac{\Delta(\ell)^2}{D(\ell)^2}.
\eeq
Before we discuss the renormalization group equations
for the scale dependent diffusion coefficient $D(\ell)$
and force correlation $\Delta(\ell)$ we notice immediately
that for L\'{e}vy flights with $\mu=f<2$ there is no correction
to the anomalous term $Dk^\mu$ in the Fokker Planck equation.
The perturbative contribution comes with a leading power
$k^2$ which is subdominant compared with the L\'{e}vy term
$k^\mu$. In the case of ordinary Brownian motion for $\mu=2$
there is a correction and we recover the results in
refs. \cite{fisher1}.

Proceeding with the discussion of Eqs. (19) and (20) we fix
as usual the dynamic exponent $z(\ell)=\mu$ such that the
diffusion coefficient stays constant under renormalization,
i.e., $D(\ell)=D$. The equation for $\Delta(\ell)$ then
takes the form
\beq
\frac{d\Delta(\ell)}{d\ell}=(2\mu-d-2)\Delta(\ell)
-B\frac{\Delta(\ell)^2}{D^2}.
\eeq
For $d$ greater that the critical dimension $d_c=2\mu-2$ Eq. (21)
has the trivial fixed point $\Delta^\ast=0$ and the quenched
disorder is {\em irrelevant}. For $d<d_c$ we obtain the non
trivial fixed point $\Delta^\ast=(d_c-d)D^2/B$ and the quenched
disorder becomes {\em relevant}.
We also note that unlike the case of Brownian motion the critical
dimension $d_c=2\mu-2$ is less than the fractal dimension $d_F=\mu$.

In order to derive the scaling properties of the force averaged
distribution \linebreak $\langle P(\k,\om)\rangle_F$ and
the means square displacement
$\langle\langle r^2(t)\rangle\rangle_F$ we use the methods discussed in refs.
\cite{aronovitz,forster1,forster2}. From the derivation
of the renormalization group equations we infer the scaling relation
\beq
\langle P(\k,\om,\Delta)\rangle_F=
e^{\alpha(\ell)}
\langle P(\k e^{\ell},\om e^{\alpha(\ell)},\Delta(\ell))\rangle_F.
\eeq
In the vicinity of either the trivial fixed point $\Delta^\ast=0$
for $d>d_c=2\mu-2$ or the fixed point $\Delta^\ast=
(d_c-d)D^2/B$ for $d<d_c$ we have,
setting $\alpha(\ell)\propto\mu\ell$,
\beq
\langle P(\k,\om,\Delta)\rangle_F=
e^{\mu\ell}
\langle P(\k e^{\ell},\om e^{-\mu\ell},\Delta^\ast)\rangle_F
\eeq
Choosing $ke^{\ell}\sim 1$ we obtain the scaling form
\beq
\langle P(\k,\om,\Delta)\rangle_F=k^{-\mu}L(k/\om^{1/\mu}),
\eeq
where L is a scaling function.
{}From Eq. (24) follows directly
\beq
\langle\langle r^2(t)\rangle\rangle_F\propto t^{2/\mu}=t^{2/z}.
\eeq
Since the scaling analysis in the L\'{e}vy case is quite similar
to the Brownian case discussed in refs. \cite{fisher1,aronovitz,fisher2}
we shall not repeat this analysis here but simply discuss and summarize
our results below.

For L\'{e}vy flights in  random quenched environment we have demonstrated
that the dynamic exponent $z$ locks onto the scaling index $\mu$,
depending on the L\'{e}vy step index $f$, independent of the presence
of weak quenched disorder. The long range superdiffusive behaviour
characteristic of L\'{e}vy flights enables the walker to escape
the inhomogeneous pinning environment and the long time behaviour
is the same as in the pure case. This result is in contrast
to the case of a Brownian walk where the dynamic exponent
$z>2$ below $d=2$, corresponding to subdiffusive behaviour.
We have also identified a critical dimension $d_c=2\mu-2$,
depending on the scaling exponent $\mu$. Below the critical
dimension the weak disorder becomes {\em relevant} as shown
by the emergence of a non trivial fixed point.

Bouchaud \cite{bouch2} has given a heuristic argument
yielding the critical dimension $d_c=\mu$ in the L\'{e}vy case.
This result is at variance with the critical dimension
$d_c=2\mu-2$ given here based on a renormalization group
analysis. It would be of interest to construe a qualitative
heuristic argument for the critical dimension $d_c=2\mu-2$ given here
and the insensitivity of the dynamic exponent $z=\mu$ to the
weak quenched disorder.

Below we have in Figure 1 plotted the scaling exponent $\mu$
and the dynamic exponent $z$ as function of the L\'{e}vy
step index $f$. In Figure 2 we have plotted the critical
dimension $d_c$ as a function of $\mu$.

\setlength{\unitlength}{1cm}

\begin{picture}(12,8)
\put(2,2){\line(0,1){4}}
\put(2,2){\line(1,0){7}}
\put(4,2){\line(0,1){0.2}}
\put(2,4){\line(1,0){0.2}}
\thicklines
\put(2,2){\line(1,1){2}}
\put(4,4){\line(1,0){3}}
\put(9.5,2){\makebox(0,0){f}}
\put(2,6.5){\makebox(0,0){$\mu$,z\, ($z=\mu$)}}
\put(4,1.5){\makebox(0,0){2}}
\put(1.5,4){\makebox(0,0){2}}
\end{picture}

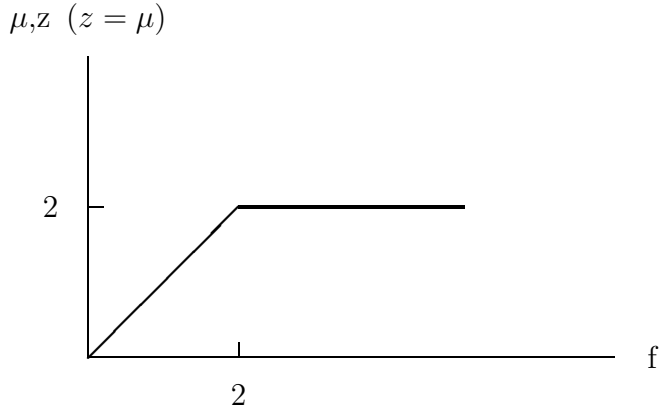
Fig.1. Plot of the scaling index $\mu$ and the dynamic exponent
$z$ as functions of the L\'{e}vy step index $f$. For $f>2$ we
have normal Brownian diffusion and we obtain corrections to $z$ below
$d=2$; for $0<f<2$ we have anomalous
L\'{e}vy superdiffusion.

\begin{picture}(12,8)
\put(2,2){\line(0,1){5}}
\put(2,2){\line(1,0){7}}
\put(4,2){\line(0,1){0.2}}
\put(6,2){\line(0,1){0.2}}
\put(2,6){\line(1,0){0.2}}
\thicklines
\put(4,2){\line(1,2){2}}
\put(9.5,2){\makebox(0,0){$\mu$}}
\put(2,7.5){\makebox(0,0){$d_c$}}
\put(4,1.5){\makebox(0,0){1}}
\put(6,1.5){\makebox(0,0){2}}
\put(1.5,6){\makebox(0,0){2}}
\end{picture}

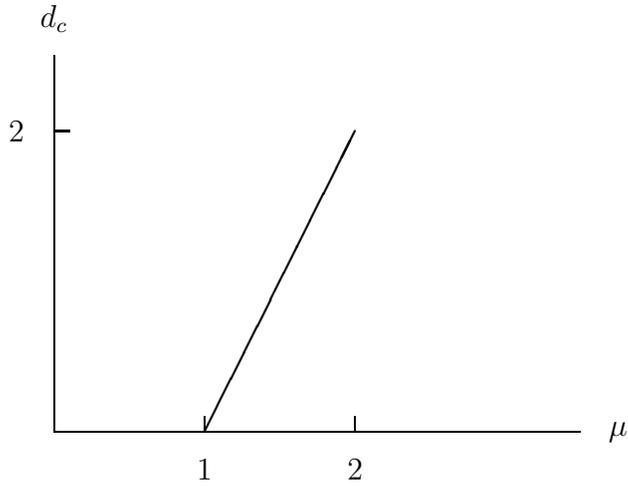
Fig.2. Plot of the critical dimension $d_c$ as a function
of the scaling index $\mu$. For $\mu=2$ we have the Brownian
case $d_c=2$; for $1<\mu<2$ $d_c$ depends linearly on $\mu$. Note that
$d_c=0$ in the ballistic case for $\mu=1$.

The author wishes to thank A. Svane, L. Mikheev, and K. B\ae kgaard
Lauritsen for helpful discussions.
This work was supported by the Danish Research Council,
grant no. 11-9001.

\end{document}